# Universal Scaling of Acoustic and Thermoacoustic Waves in Compressible Fluids

**Mario Tindaro Migliorino**[1]†, **Carlo Scalo**[1]

[1]Department of Mechanical Engineering, Purdue University, West Lafayette, IN 47907, USA

We have derived the set of reference scaling parameters yielding collapse of isentropic acoustic and thermoacoustic (or heat-release-induced) waves across different pure compressible fluids with an assigned equation of state. The resulting reference pressure and velocity are consistent with classic acoustic scaling. The reference temperature and heat release rate need to be expressed in terms of the isobaric thermal expansion coefficient $\alpha^*_{p_0}$ to ensure collapse of all thermo-fluid-dynamic fluctuations across all fluids. The proposed scaling is extended to non-isentropic waves and verified against data from highly-resolved one-dimensional Navier-Stokes simulations. Conditions tested include freely propagating isentropic acoustic waves and thermoacoustic compression waves up to shock strength of 4.27, for six different supercritical fluids in states ranging from compressible liquid to near-ideal gas, and spanning seven orders of magnitude of imposed heat release rate.

## 1. Introduction

Waves in compressible fluids are propagating disturbances affecting all thermo-fluid-dynamic variables. However, common practice in acoustics is to use, when possible, pressure and velocity fluctuations (or pressure only) as the sole working variables (Lighthill 1952). This choice is consistent with the fundamental nature of sound waves, i.e. self-propagating patterns of compressions and dilatations, inducing and induced by spatial gradients in particle displacements or velocities. No other variables but pressure and velocity are thus needed to intuitively understand sound propagation and the mechanical or acoustic power associated with it.

As a result, the traditional approach to deriving a dimensionless set of linearized governing wave equations has been focused on collapsing pressure and velocity fluctuations only, with little consideration given to other fluctuations, such as temperature, enthalpy or internal energy, which are still present and equally relevant (Doak 1998). For example, temperature fluctuations are of primary importance in studies of thermophones (Suk *et al.* 2012), and thermoacoustic (or heat-release-induced) waves (Taylor 1950; Kassoy 1979; Clarke *et al.* 1984; Brown & Churchill 1995; Carlès 2006; Shen & Zhang 2011), to cite a few. Moreover, thermoacoustic-wave-induced temperature fluctuations are the governing mechanism for thermal relaxation in near-critical fluids in enclosed cavities (Onuki *et al.* 1990; Boukari *et al.* 1990; Zappoli *et al.* 1990; Amiroudine & Zappoli 2003; Zappoli 2003; Onuki 2007; Carlès 2010), referred to as the Piston Effect.

The identification of the correct set of scaling parameters is necessary to derive a dimensionless set of governing equations with universal applicability. To the authors' knowledge, there have been no prior attempts towards developing an appropriate dimensionless scaling strategy able to provide a unified description of linear waves across different fluids and in different states. For example, as discussed in section 2 and shown in figure 2, simply using the fluid's base temperature to scale temperature fluctuations (Rienstra & Hirschberg 2016), does *not* yield dimensionless collapse of temperature fluctuations, even in the simple case of ideal gases.

† Email address for correspondence: migliom@purdue.edu

| state | $p_0^*/p_{cr}^*$ | $T_0^*/T_{cr}^*$ | $CO_2$ | $O_2$ | $N_2$ | $CH_3OH$ | R-134a | R-218 |
|---|---|---|---|---|---|---|---|---|
| PL | 1.1 | 0.89 | ● | ▲ | ▼ | ■ | ◆ | ⬟ |
| PB | 1.1 | 1.02 | ◉ | ▲ | ▼ | ▪ | ◆ | ⬟ |
| PG | 1.1 | 1.11 | ○ | △ | ▽ | ▫ | ◇ | ⬠ |
| IG | 1.1 | 2.20 | ○ | △ | ▽ | □ | ◇ | ⬠ |
| $p_{cr}^*$ (MPa) | | | 7.3773 | 5.043 | 3.398 | 8.097 | 4.059 | 2.68 |
| $T_{cr}^*$ (K) | | | 304.13 | 154.58 | 126.2 | 512.64 | 374.26 | 345.1 |

TABLE 1. Marker legend for six fluids each considered in four different reference states indicated with the subscript '0'. Greyscale levels in the markers range from black: heavy fluid, pseudo-liquid (PL); to white: light fluid, near ideal-gas (IG). Such reference conditions are used as the acoustic base state in section 2 and initial conditions in section 3; $p_{cr}^*$ and $T_{cr}^*$ are the dimensional values of fluid-specific critical pressures and temperatures.

In sections 2 and 3.1 we address this knowledge gap by deriving a set of reference scaling parameters for all thermo-fluid-dynamic fluctuations yielding full collapse of isentropic acoustic and thermoacoustic planar waves. In section 3.2 we extend the derived scaling to pressure jumps across thermoacoustic shock waves up to shock strength of 4.27.

To verify the proposed scaling, highly-resolved single-phase high-order fully compressible Navier-Stokes simulations are performed with the solver *Hybrid* (Larsson & Lele 2009) for six different fluids at supercritical pressure conditions $p_0^* = 1.1 \, p_{cr}^*$ (table 1). The base temperature $T_0^*$ is varied to achieve a range of conditions including: pseudo-liquid (PL), pseudo-boiling (PB), pseudo-gaseous (PG), modeled with the Peng-Robinson (Peng & Robinson 1976) equation of state (EoS) and Chung's method (Chung *et al.* 1988; Poling *et al.* 2001); and near-ideal gas (IG), modeled as a perfect gas and with Sutherland's law. This manuscript contains data from a total of 187 numerical simulations.

## 2. Scaling of Isentropic Acoustic Waves

We start from the simple case of planar isentropic acoustic waves evolving in a uniform quiescent base state, hereafter indicated with the subscript '0'. All dimensional quantities are denoted by the superscript ($*$), which is omitted in their dimensionless counterpart. In this case pressure $\delta p^*$ and velocity $\delta u^*$ fluctuations are the only quantities needed to completely characterize the flow; the proper and commonly used normalization choice is

$$\delta p = \frac{\delta p^*}{\rho_0^* a_0^{*2}}, \qquad \delta u = \frac{\delta u^*}{a_0^*}, \qquad x = \frac{x^*}{\ell^*}, \qquad t = \frac{t^*}{\ell^*/a_0^*}, \qquad (2.1)$$

where $\rho_0^*$ and $a_0^* = \sqrt{\partial p^*/\partial \rho^*|_{s^*,0}}$ are the base density and isentropic speed of sound, and $x^*$ and $t^*$ the independent spatial and temporal coordinates, and $\ell^*$ a reference length scale. Applying the normalization in Eq. (2.1) to the linearized continuity and

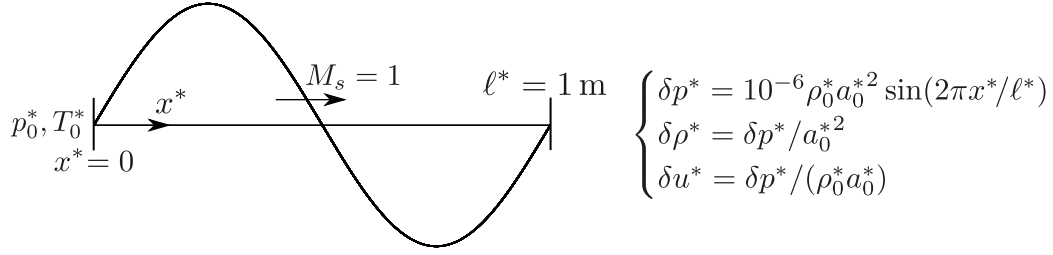

FIGURE 1. Initial conditions for isentropic right-traveling acoustic wave in a periodic domain $[0, \ell^*]$. The initial fluctuating temperature field is obtained via the assigned equation of state.

momentum equations, assuming isentropic flow, yields

$$\frac{\partial}{\partial t}\delta u = -\frac{\partial}{\partial x}\delta p, \qquad \frac{\partial}{\partial t}\delta p = -\frac{\partial}{\partial x}\delta u, \tag{2.2}$$

whose solution in an unbounded domain can be expressed without loss of generality in the self-similar form

$$\delta u_\pm = \pm \delta p_\pm = f_\pm(\xi_\pm). \tag{2.3}$$

The two functions $f_\pm(\cdot)$ of the traveling-wave coordinate $\xi_\pm = x \mp M_s t$ (where $M_s = a_0$ is the wave Mach number, unitary in this case) can be independently and arbitrarily assigned with the caveat that $\max_\xi |f(\xi)| \ll 1$ to honor assumptions of linearity. The dependency on the base state, and hence also on the specific fluid properties, of the governing equations (2.2) has been completely absorbed by the normalization (2.1). The steps leading to Eq. (2.2) and Eq. (2.3) do *not* require the specification of an equation of state nor of an explicit normalization for temperature fluctuations $\delta T^*$; however, they do entail the normalization $\delta \rho = \delta \rho^* / \rho_0^*$. Without loss of generality, numerical verification of the proposed scaling will focus on right-traveling waves (figure 1).

We now observe how the commonly adopted normalization of temperature fluctuations $\delta T^*$, using the base state temperature $T_0^*$ as a reference, does not collapse isentropic temperature perturbations, associated to the same acoustically scaled waveform $f(\xi)$ (figure 2a), across different fluids, even in ideal gas conditions (figure 2b). This issue is resolved by noticing that an isentropic fluctuation of a generic quantity $\delta \varphi^*$ can be expressed as a sole function of pressure fluctuations via evaluation of the thermodynamic derivative $\delta \varphi^* / \delta p^* = \partial \varphi^* / \partial p^* |_{s^*, 0}$, yielding

$$\delta T^* = \frac{\alpha_{p_0}^* T_0^*}{\rho_0^* c_{p_0}^*}\delta p^*, \qquad \delta h^* = \frac{1}{\rho_0^*}\delta p^*, \qquad \delta e^* = \frac{p_0^*}{\rho_0^{*2} a_0^{*2}}\delta p^*, \tag{2.4}$$

where $\delta h^*$ and $\delta e^*$ are the specific (per unit mass) enthalpy and internal energy fluctuations, $c_{p_0}^*$ is the isobaric specific thermal capacity and

$$\alpha_{p_0}^* = -\rho_0^{*-1}\,\partial \rho^* / \partial T^*|_{p^*, 0} \tag{2.5}$$

is the isobaric thermal expansion coefficient, both calculated at base state conditions. Applying the relation $a_0^{*2} T_0^* \alpha_{p_0}^{*2} / c_{p_0}^* = \gamma_0 - 1$, where $\gamma_0 = c_{p_0}^* / c_{v_0}^*$ is the ratio of specific isobaric and isochoric thermal capacities, and the normalization in Eq. (2.1), to Eq. (2.4), the correct normalization achieving the desired collapse (figure 2d) reads

$$\delta T = \frac{\alpha_{p_0}^*}{\gamma_0 - 1}\delta T^*, \qquad \delta h = \frac{1}{a_0^{*2}}\delta h^*, \qquad \delta e = \frac{\widetilde{\gamma}_0}{a_0^{*2}}\delta e^*, \tag{2.6}$$

where $\widetilde{\gamma}_0 = \rho_0^* a_0^{*2} / p_0^*$ is the isentropic exponent (Iberall 1948).

Equations (2.1) and (2.6) define the complete set of reference scaling parameters,

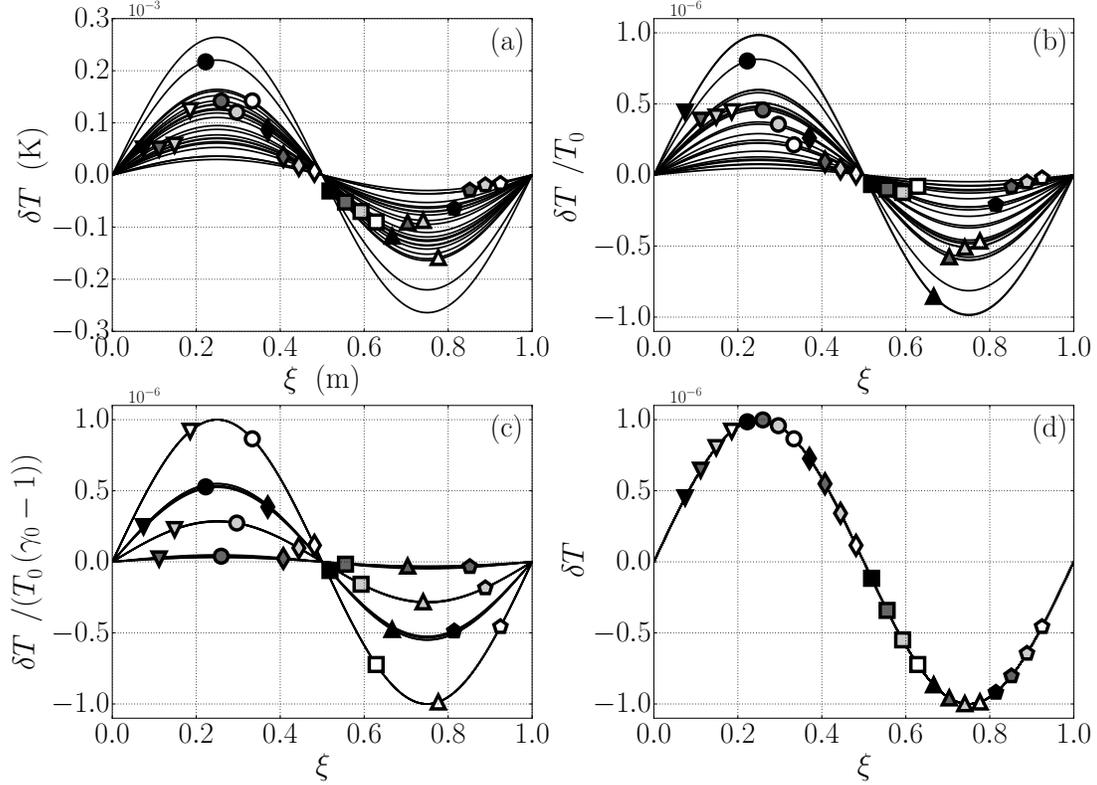

FIGURE 2. Temperature fluctuations from inviscid computations of right-traveling acoustic waves for $f(\xi) = 10^{-6}\sin(\xi)$ (see Eq. (2.3)) for all fluids and conditions in table 1. (a) Dimensional temperature perturbations; (b) scaling based on base temperature; (c) universal scaling only valid for IG; (d) universal scaling valid for all fluids (table 2). Same results are obtained for other variables such as $\delta e$, $\delta h$, and $\delta \rho$ (not shown).

| $\delta p^*_{ref}$ | $\delta u^*_{ref}$ | $\delta \rho^*_{ref}$ | $\delta T^*_{ref}$ | $\delta h^*_{ref}$ | $\delta e^*_{ref}$ | $\Omega^*_{ref}$ |
|---|---|---|---|---|---|---|
| $\rho_0^* a_0^{*2}$ | $a_0^*$ | $\rho_0^*$ | $(\gamma_0 - 1)/\alpha^*_{p_0}$ | $a_0^{*2}$ | $a_0^{*2}/\widetilde{\gamma}_0$ | $2\rho_0^* a_0^* c^*_{p_0}/\alpha^*_{p_0}$ |

TABLE 2. Set of reference scaling parameters yielding universal collapse of isentropic acoustic and thermoacoustic waves across different fluids.

summarized in table 2, collapsing all isentropic thermo-fluid-dynamic fluctuations across different fluids, and, incidentally, also among themselves,

$$\delta p = \delta \rho = \delta T = \delta h = \delta e = f(\xi), \tag{2.7}$$

which is a direct result of the single degree of thermodynamic freedom. Moreover, $\delta u = \delta p$ and $\delta u = -\delta p$ for right- and left-traveling waves, respectively.

Normalizing temperature, internal energy, and enthalpy fluctuations using the base temperature $T_0^*$ only yields collapse across different ideal gases with the same value of $\gamma_0$, since in that case the proposed scaling parameters revert to $\delta T^*_{ref}|_{IG} = (\gamma_0 - 1)T_0^*$, $\delta e^*_{ref}|_{IG} = R^*T_0^*$, $\delta h^*_{ref}|_{IG} = \gamma_0 R^*T_0^*$, where $R^*$ is the gas constant. Ideal gas temperature perturbations made dimensionless only via $T_0^*$ (figure 2b), in fact, do not collapse unless $\gamma_0 - 1$ is also taken into consideration (figure 2c). Finally, full collapse across all fluids and for all conditions (figure 2d) can only be achieved employing the correct reference temperature $\delta T^*_{ref} = (\gamma_0 - 1)/\alpha^*_{p_0}$ (figure 3). The same result is obtained for enthalpy and internal energy fluctuations (not shown) by using their respective scaling parameters indicated in table 2.

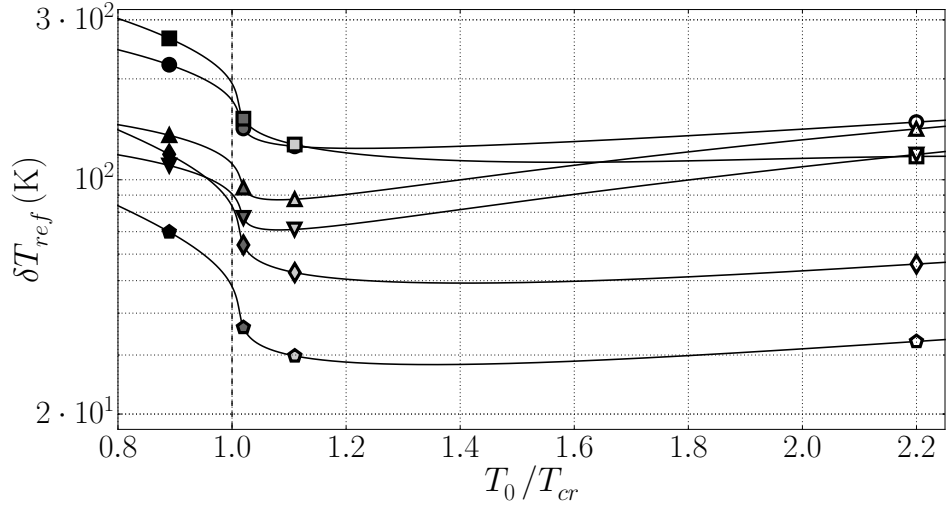

FIGURE 3. Reference scaling parameter for temperature fluctuations, $\delta T^*_{ref}$, versus reduced temperature for all fluids and conditions in table 1.

## 3. Scaling of Thermoacoustic (or Heat-Release-Induced) Waves

In the second part of this paper we extend the scaling strategy discussed above to thermoacoustic waves ranging in intensity from isentropic compressions to shock waves; the latter are considered only during their stage of *equilibrium inviscid propagation*, that is right after shock formation (i.e. full coalescence of compression waves) but before the onset of viscous decay, during which jumps of thermo-fluid-dynamic variables are constant and independent from viscosity. We rely on fully resolved computations of the governing equations for mass and momentum, respectively, and total energy,

$$\frac{\partial \rho^*}{\partial t^*} + \frac{\partial \rho^* u^*}{\partial x^*} = 0, \qquad \frac{\partial \rho^* u^*}{\partial t^*} + \frac{\partial (\rho^* u^{*2} + p^*)}{\partial x^*} = \frac{\partial \tau^*}{\partial x^*}, \tag{3.1}$$

$$\frac{\partial \rho^* E^*}{\partial t^*} + \frac{\partial (\rho^* u^* E^* + p^* u^*)}{\partial x^*} = \frac{\partial u^* \tau^*}{\partial x^*} - \frac{\partial q^*}{\partial x^*} + \dot{Q}^*, \tag{3.2}$$

where the Newtonian viscous stress $\tau^* = (4/3)\mu^* \partial u^*/\partial x^*$ is expressed according to Stokes's hypothesis, the heat flux $q^* = -k^* \partial T^*/\partial x^*$ is modeled with Fourier heat conduction, $\mu^* = \mu^*(\rho^*, T^*)$ is the dynamic viscosity and $k^* = k^*(\rho^*, T^*)$ is the thermal conductivity. Finally, $\dot{Q}^*$ (W/m$^3$) is the volumetric heat release rate, expressed as

$$\dot{Q}^*(x^*, t^*) = \Omega^* g^*(x^*), \qquad g^*(x^*) = \frac{1}{\ell^* \sqrt{2\pi}} e^{-\frac{1}{2}(x^*/\ell^*)^2}, \tag{3.3}$$

where $\Omega^*$ (W/m$^2$) is the planar heat release rate and $g^*$ (m$^{-1}$) is a Gaussian function with unitary (non-dimensional) integral on the real axis, with characteristic width $\ell^* = 0.75\,\mu$m, inspired by the experiments of Miura *et al.* (2006). Simulations are carried out only for $x^* \geqslant 0$, with adiabatic wall conditions imposed at $x^* = 0$, and halted before perturbations reach the right boundary (located at $x^* = 40\mu$m). The computational domain is sufficiently long to allow shocks to reach conditions of equilibrium inviscid propagation. The shocks are resolved with a minimum number of 14 grid points, corresponding to a grid spacing of $\Delta x^* = 0.04\,\mu$m, and with a maximum time step of $\Delta t^* = 0.01$ ns. Simulations are performed for values of $\Omega^*$ reported in table 3 and all initial conditions in table 1.

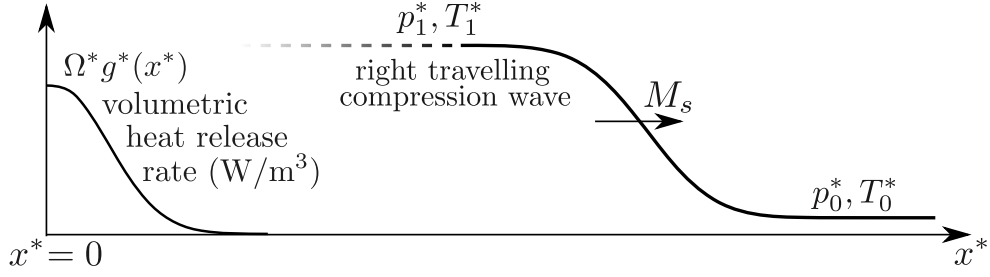

FIGURE 4. Computational setup for thermoacoustic wave generation.

| $\Omega^*$ (W/m$^2$) | $10^5$ | $10^7$ | $10^9$ | $(1,3,6) \times 10^{10}$ | $^\dagger (1,3,6) \times 10^{11}$ | $^\ddagger 10^{12}$ |
|---|---|---|---|---|---|---|

TABLE 3. Planar heat release rates used in numerical simulations of thermoacoustic waves. Values of the order of $10^{11}$ ($^\dagger$) are used only for PL conditions, and of $10^{12}$ ($^\ddagger$) only for CH$_3$OH in PL conditions (see table 1).

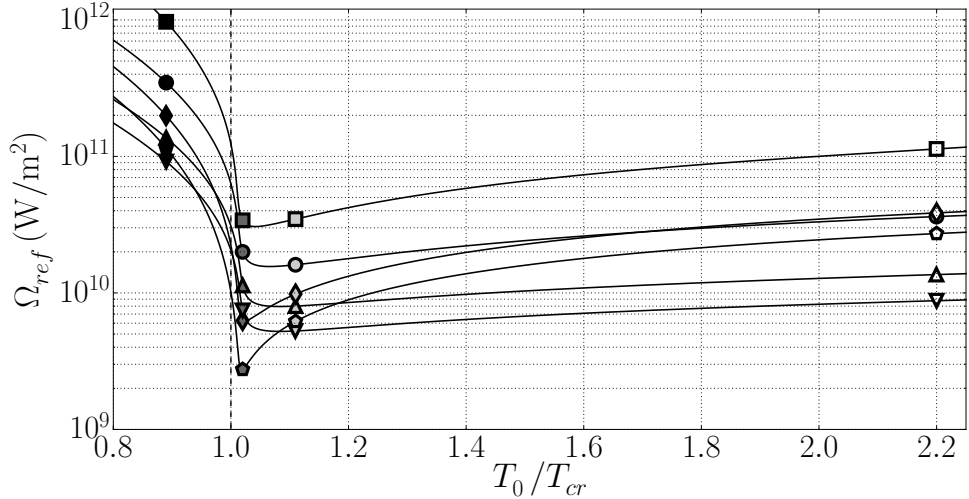

FIGURE 5. Reference scaling parameter for heat release rate, $\Omega^*_{ref}$, versus reduced temperature for all fluids and conditions in table 1.

### 3.1. *Near-isentropic Thermoacoustic Compression Waves*

Building upon Miura *et al.* (2006)'s derivation, for low heat release rates (quasi-isentropic regime), the amplitude of the generated compression waves is predicted via

$$\Pi^* = p_1^* - p_0^* = \frac{a_0^* \alpha_{p_0}^*}{c_{p_0}^*} \frac{\Omega^*}{2}, \qquad (3.4)$$

where the subscript '1' indicates the post-compression state (figure 4). Eq. (3.4) should be made dimensionless as follows:

$$\Pi = \frac{\Pi^*}{\delta p_{ref}^*} = \frac{\Pi^*}{\rho_0^* a_0^{*2}} = \frac{\Omega^*}{2\rho_0^* a_0^* c_{p_0}^*/\alpha_{p_0}^*} = \frac{\Omega^*}{\Omega_{ref}^*} = \Omega, \qquad (3.5)$$

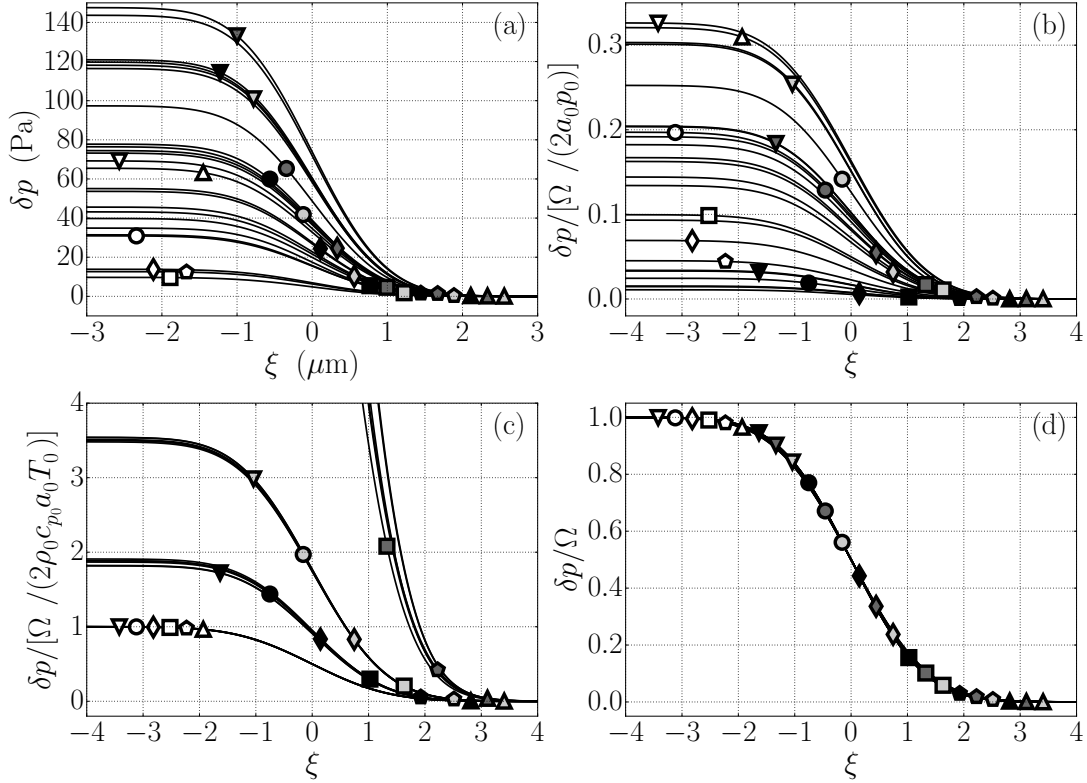

FIGURE 6. Pressure fluctuations of quasi-isentropic thermoacoustic waves for dimensional heat release rate $\Omega^* = 10^5 \, \text{W/m}^2$. (a) Dimensional pressure profiles; (b) scaling proposed by Chu (1955); (c) scaling only valid for ideal gases; (d) correct universal scaling (Eq. 3.5, table 2).

where $\Omega^*$ is normalized with $\Omega^*_{ref}$ (figure 5, table 2), and $\Pi$ is the shock strength (Thompson 1971), with $\Pi << 1$ for quasi-isentropic compressions. Under such conditions, all the thermo-fluid-dynamic jumps are only a function of $\Pi$, consistently with Eq. (2.7).

Analogously to figure 2, figure 6 shows how previously adopted normalizations of heat release rate for thermoacoustic waves (Chu 1955) lead to no (figure 6b) or incomplete (figure 6c) collapse, ultimately only achieved in full by the proposed scaling (figure 6d).

In the near-isentropic low-wave-amplitude regime ($\Omega < 10^{-1}$), Eq. (3.5) yields full collapse of the numerical data across all fluids and conditions (figure 7) confirming the universality of the proposed scaling. While the extraction of pressure jumps $\Pi^*$ from the numerical simulation data is trivial in such regime due to the flatness of the post-compression state (figure 6), for larger wave amplitudes the jump is evaluated during the inviscid propagation phase of the shock. As discussed below, the non-isentropic nature of wave propagation in this regime will naturally require the introduction of another parameter to account for the second degree of thermodynamic freedom.

### 3.2. *Non-Isentropic Thermoacoustic Waves*

Thermoacoustic wave amplitudes depart from the prediction in Eq. (3.5) for $\Omega > 10^{-1}$ (see insets of figure 7) following a fluid-specific and state-specific law, which appears to be functionally identical to the one valid for ideal gases. In the latter case an exact parametrization has been derived by Chu (1955)[pp. 20-26]. Based on this observation, we speculate that shock-amplitude prediction is still possible in this regime by heuristically extending Chu's parametrization via an effective ratio of specific heats $\gamma_\pi$,

$$\Omega = \sqrt{2}\Pi(\gamma_\pi \Pi + 1)\left(2 + (\gamma_\pi + 1)\Pi\right)^{-0.5}, \quad (3.6)$$

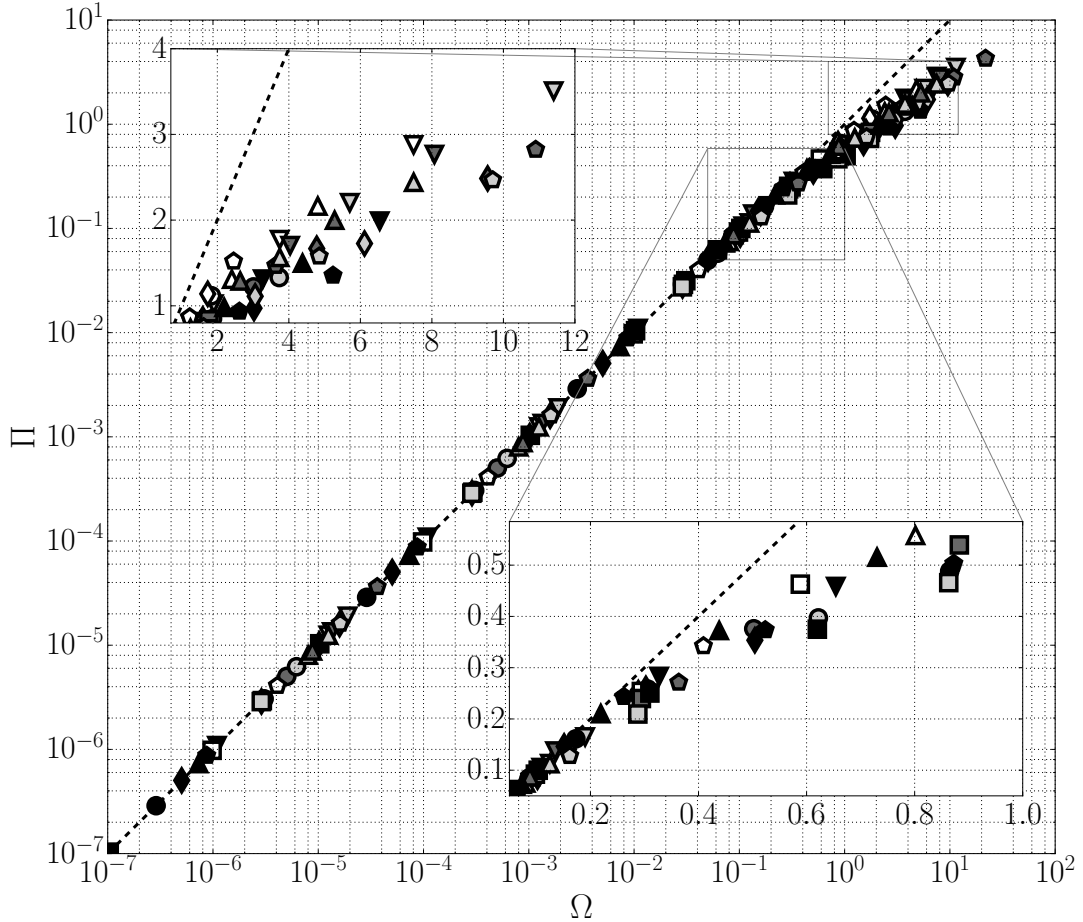

FIGURE 7. Shock strength versus dimensionless heat release rate from numerical computations (symbols, tables 1 and 3). The quasi-isentropic prediction of Eq. (3.5) is the dashed line.

| $\gamma_\pi$ | $CO_2$ | $O_2$ | $N_2$ | $CH_3OH$ | R-134a | R-218 |
|---|---|---|---|---|---|---|
| PL | 3.7748 | 3.1987 | 3.1464 | 3.8536 | 5.0326 | 5.6114 |
| PB | 2.7752 | 2.1436 | 2.0418 | 2.7796 | 3.4312 | 3.6123 |
| PG | 3.3653 | 2.4135 | 2.0356 | 3.9855 | 4.1466 | 3.9855 |
| IG | 1.2455 | 1.4722 | 1.4843 | 1.1106 | 1.0743 | 1.0474 |

TABLE 4. Values of $\gamma_\pi$ obtained by least-square fit of the numerical data in figure 8.

where, for IG conditions, $\gamma_\pi = \gamma_0$, hence recovering the original formula by Chu. Eq. (3.6) reverts to Eq. (3.5) for $\Pi \to 0$, removing the dependency from $\gamma_\pi$, consistently with the single degree of thermodynamic freedom intrinsic to isentropic or near-isentropic waves. When non-reversible entropy generation occurs across the thermoacoustic compression, the specification of the fluid-specific and state-specific $\gamma_\pi$ (second degree of thermodynamic freedom) is necessary to predict the pressure jump. Values of $\gamma_\pi$ have been fitted against the data from the numerical simulations via Eq. (3.6) (table 4) demonstrating remarkable fit quality (figure 8) up to shock strengths of $\Pi = 4.27$.

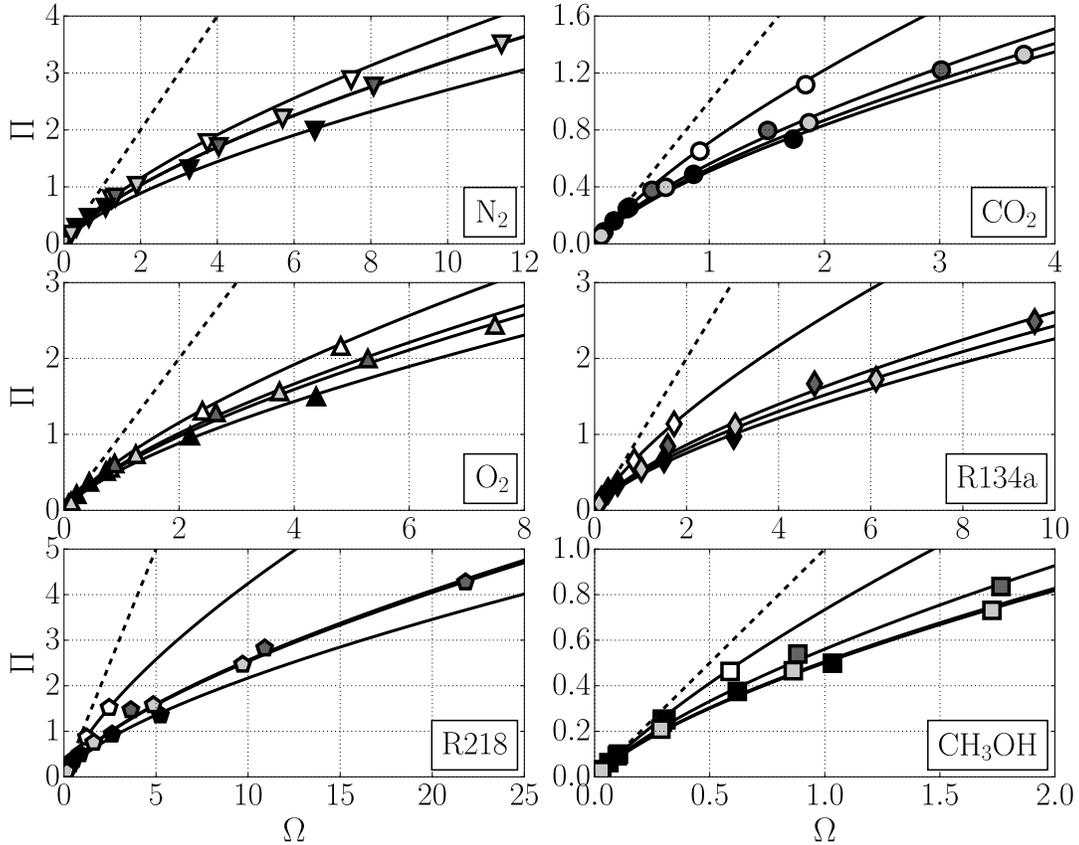

FIGURE 8. Fluid-by-fluid shock strengths for high heat release rates: numerical computations (symbols) are fitted with Eq. (3.6) (solid lines), resulting in $\gamma_\pi$ values of table 4 and deviating from the quasi-isentropic prediction of Eq. (3.5) (dashed line).

## 4. Conclusions

In summary, we have derived the correct set of reference scaling parameters able to achieve collapse of isentropic acoustic and thermoacoustic waves across different pure compressible fluids in conditions ranging from liquid-like to gas-like. Data from highly resolved one-dimensional Navier-Stokes numerical simulations has been adopted to verify the effectiveness of the proposed scaling strategy and to aid its extension to the non-isentropic regime, where thermoacoustic shock waves have been investigated. We have demonstrated that in this regime a new effective ratio of specific heats, $\gamma_\pi$, can be specified to enable prediction of shock strengths.

## Acknowledgements

Mario Tindaro Migliorino acknowledges the support of the Frederick N. Andrews and Rolls-Royce Doctoral Fellowships at Purdue University. The authors thank Mr. Pat Sweeney (Rolls-Royce) and Prof. Stephen D. Heister (Purdue) for the fruitful discussions that have inspired the need for scaling across different fluids. The computing resources were provided by the Rosen Center for Advanced Computing (RCAC) at Purdue University and Information Technology at Purdue (ITaP).